\documentclass[a4paper]{VisionStyle}
\usepackage{epsfig}

\begin{document}

\title{Spectral properties of faint hard X-ray selected sources}

\author{E.\,Piconcelli\inst{1,2} \and L.\,Bassani\inst{1}  \and M.\,Cappi\inst{1} \and F.\,Fiore\inst{3} \and G.\,DiCocco\inst{1}
\and M.\,Trifoglio\inst{1}} 

\institute{
  IASF/CNR, Via Gobetti 101, Bologna, I-40129, Italy 
\and 
  Dipartimento di Astronomia, Via Ranzani 1, Bologna, I-40127, Italy
 \and 
  Osservatorio Astronomico di Roma, Via Frascati 33, Monteporzio, I-00044, Italy}

\maketitle 

\begin{abstract}
We present preliminary results from a XMM-Newton spectroscopic survey of hard (2-10 keV) X-ray sources detected serendipitously
in seven EPIC FOVs. All observations were performed during the performance/verification phase.
The resulting sample consists of 39 sources with fluxes ranging from 3 to 80 $\times$ 10$^{-14}$  erg cm$^{-2}$ s$^{-1}$ in the 2-10 keV 
band. Most of these sources (23 out of 39, i.e. 60\%) are fainter than 8 $\times$ 10$^{-14}$ erg cm$^{-2}$ s$^{-1}$.
The sources have been selected in order to obtain detailed X-ray spectral information, that are unprecedented at such low flux 
levels and for such a large sample. 
 To date we also have optical counterparts for
23 out of 39 objects, with 0.0058 $\leq$ $z$ $\leq$ 1.187. 

About 70\% of the sources have spectra well fitted with a simple power law model with Galactic absorption. For 9 objects (five of type 1, one  of type 2 and  three normal galaxies) absorption in excess to the Galactic one is required. Additional emission/absorption components (i.e. soft excess, warm absorber) are observed in 4 sources.

\keywords{galaxies: active -- X-rays: galaxies -- cosmic X-ray background}
\end{abstract}

\section{Scientific goal} 
The unprecedented sensitivity and excellent imaging capabilities of XMM-Newton and Chandra observatories provide an accurate
 and fascinating view of the whole population of cosmic X-ray sources, allowing to investigate a very large, 
and previously unexplored, parameters space.

Recent deep observations with Chandra (Tozzi et al. 2001; Brandt et al. 2001; Mushotzky et al. 2000) and 
XMM-Newton (Hasinger et al. 2001) in the range between 0.1 to 10 keV have in fact definitively probed that 
the bulk of the cosmic X-ray background (CXB) is due to the emission of discrete sources
along the cosmic time.
Such deep surveys allowed to reach extremely low flux levels ($\sim$6 $\times$ 
10$^{-17}$ erg cm$^{-2}$ s$^{-1}$ and $\sim$5 $\times$ 10$^{-16}$ erg cm$^{-2}$ s$^{-1}$ between 0.5-2 keV and 2-10 keV, respectively) and to resolve $\sim$80-90\% of the CXB below 10 keV.
Multiwavelength follow-up observations of these X-ray surveys indicate that the majority of the X-ray sources are associated with active 
galactic nuclei (AGNs) while the remaining ones are bright early-type galaxies or optically faint ($I >$ 24) objects (likely obscured AGNs).

Current CXB synthesis models (Madau, Ghisellini \& Fabian 1994; Comastri et al. 1995) are based on a mixture of type 1 
(unabsorbed) and type 2 (absorbed) objects. They predict the presence of a substantial number of high luminosities
 (log$L_{X} >$ 44 erg s$^{-1}$) heavily obscured (log$N_{H} >$ 23 cm$^{-2}$) sources, but only a handful of convincing cases
of pure reflection type 2 QSOs have been discovered so far (see i.e. Norman et al. 2001; Franceschini et al. 2000). 
Moreover recent works reveal the existence of previously undetected intriguing sources, most of which are X-ray faint, 
i.e. broad line AGNs which suffer from intrinsic X-ray absorption (Maloney \& Reynolds 2001; Fiore et al. 2001; Wilkes et al. 2001) and 
X-ray loud optically quiet galaxies (Fiore et al. 2000; Allen, Di Matteo \& Fabian 2000). 
These objects show intrinsic flat X-ray slopes which match well with the CXB one 
($\Gamma$ = 1.4, Vecchi et al. 1999) so that they could be significant in explaining the flat spectral shape of the CXB.

So far, howewer, very few broadband spectra of X-ray sources with $F_{2-10} \leq$ 10$^{-13}$ erg cm$^{-2}$ s$^{-1}$ have been published in 
the literature and their properties (i.e. spectral index, absorption column density and strenght of features due to reprocessing)
have been  poorly investigated.\\ 

Such detailed spectral information are now accessible for the first time with few tens 
kilosecond observations of the EPIC detector on-board the XMM-Newton satellite. We have thus started an extensive program aimed at
performing an accurate spectroscopic analysis of hard X-ray (2-10 keV) selected serendipitous sources in a large number of EPIC fields.
We have chosen EPIC because of its unprecedented high sensitivity in the hard X-ray band and of its arcseconds 
positional accuracy both of which are crucial to carry out such kind of study.
\section{The sample}
Here we report spectral results of a sample of 39 sources observed 
in a total of seven public EPIC fields during the performance verification phase 
of XMM-Newton. These sources have been detected with a 5$\sigma$ significance in the 2-10 keV band using {\it EBOXDETECT}, 
the standard sliding cells algorithm of the {\it SAS} software package. 
The measured $F_{2-10keV}$ range from 3 to 80 $\times$ 10$^{-14}$ erg cm$^{-2}$ s$^{-1}$ with  $\geq$ 60\% of the objects showing 
a $F_{2-10keV} <$ 10$^{-13}$ erg cm$^{-2}$ s$^{-1}$.

\begin{figure}[ht]
\begin{center}
\epsfig{file=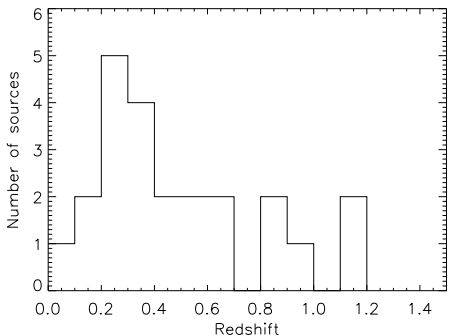,width=7cm}
\end{center}
\caption{Redshift distribution of the 23 optically identified sources in the sample. The values span from $z$ = 0.0058 to $z$ = 1.187. The
average redshift is $\langle$$z$$\rangle$ $\sim$0.5.}
\label{epiconcelli-F11_fig:fig1}
\end{figure}

\begin{figure}[ht]
\begin{center}
\epsfig{file=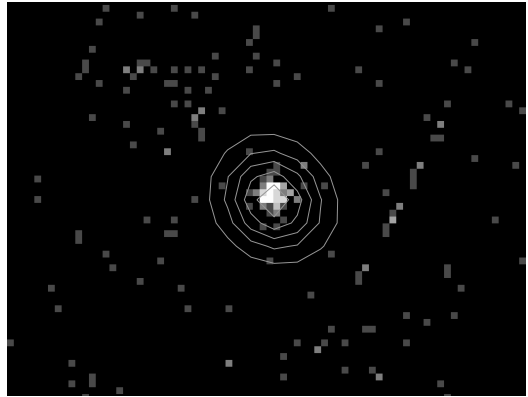,width=7cm}
\end{center}
\caption{MOS1 image of XMMU J000100.0-250459 with the superimposed contours of its radiocounterpart NVSS J000100-250503.}
\label{epiconcelli-F11_fig:fig2}
\end{figure}

23 out of 39 sources in the sample have optical identification: 13 are from the literature and the remaining 10 are from the HELLAS2XMM survey 
(Fiore et al. 2002, in preparation).  Identification of the remaining sources is 
being planned too. All these sources have $z >$ 0.1, except for the nearby normal galaxy NGC 4291 at $z$ =0.0058.
The redshift distribution over the entire sample is shown in Figure~\ref{epiconcelli-F11_fig:fig1}.
16 X-ray sources are optically classified as broad line quasars, three are normal
galaxies, two are narrow emission line galaxies, one is a Seyfert 2 galaxy and, finally, one is a  red quasar.  
We also cross-correlate our sample with the most popular radio catalogues (NVSS and FIRST) available on-line:
we find that three X-ray sources have radio counterpart, however these radio surveys do not cover the whole sky,
 so this result is not complete. As an example, the MOS1 image of the X-ray source XMMU J000100.0-250459 with the superimposed radio contours of NVSS J000100-250503 is
shown in Figure~\ref{epiconcelli-F11_fig:fig2}
Combining optical and radio information we infer that these three sources are all radio loud (RL) AGNs.


\section{The spectral properties}
To date the analysis of EPIC spectra of all our sources is still not complete hence here we report preliminar results of this work. 
The final and more detailed description of our study will be presented in a forthcoming paper (Piconcelli et al. 2002, in preparation).

\begin{figure}[ht]
\begin{center}
\epsfig{file=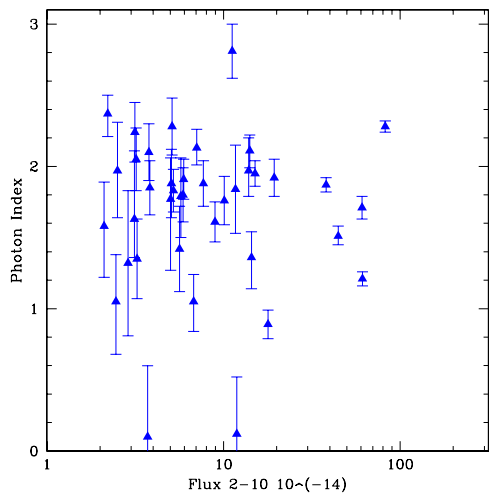,width=7cm}
\end{center}
\caption{Plot of photon index versus flux in the 2-10 keV band for the present sample, using model A.}  
\label{epiconcelli-F11_fig:fig3}
\end{figure}

We start to fit the spectra with a single power law modified by the Galactic column density (model A) in order to describe the general 
form of the continuum and to find the deviations from the model, indicative of the presence of extra features and absorption. 
In Figure~\ref{epiconcelli-F11_fig:fig3} we show the results of this spectral fitting. 
The photon spectral indeces span from 0.1 to 2.8, but most of them cluster around 1.8-2.0 e.g. the typical values found for broad 
line unabsorbed AGNs (as are the majority of the optically identified sources in the sample). 
Applying this model we find an average spectral index $\langle\Gamma\rangle$ = 1.71$\pm$0.04. 
If we consider only the 19 sources with  $F_{2-10keV} \geq$ 7 $\times$ 10$^{-14}$ erg cm$^{-2}$ s$^{-1}$ (i.e. our completeness flux limit)
we obtain a $\langle\Gamma\rangle$ = 1.68$\pm$0.04. This value is steeper than that found by \cite*{epiconcelli-F11:ueda99} and Della Ceca et al. (1999) on the basis of stacked spectral analysis of ASCA sources at a similar flux limit. On the other hand it matches well with the results
 obtained from a 130 ks Chandra
observation of the Chandra Deep Field South reported by Giacconi et al. (2001). These authors found, in fact,    
$\langle\Gamma\rangle$ = 1.71$\pm$0.07 again from a stacked spectrum analysis of sources brighter than $\sim$2 $\times$ 10$^{-14}$ erg cm$^{-2}$ s$^{-1}$.
 
It is worth noting that flat slopes ($\Gamma$ = 1.0-1.2) are present in any case;  this is probably  
indicative of a high column density which obscures the X-ray emission (absorption cross-section
 $\sigma(E) \propto$ $E^{-2.6}$). Viceversa a very steep spectral index could mask a soft excess component.

We therefore also apply more detailed fitting models to those sources showing a complex spectrum and for which model A does
 not give a satisfactory $\chi^{2}$. 
We find that 9 out of 39 sources require an additional extra absorption component. 
In all these cases we find $N_{H}$ in the range of 10$^{21-22}$ cm$^{-2}$ except for the Seyfert 2 galaxy XMMU J030911.9$-$765824, for which we obtain $N_{H} \geq$  10$^{23}$ cm$^{-2}$.
Interestingly enough, two out of three radio loud objects have an absorption column density in excess to the Galactic value. This fact was 
already found in other X-ray observations of RL AGNs (\cite{epiconcelli-F11:cappi97}; Sambruna, Eracleous \& Mushotzky 1999) and could have relevant 
implications about the cosmic evolution of this class of objects.
Furthermore the two optically dull X-ray loud galaxies (i.e. objects with no sign of nuclear activity but with unusual log$L_{X} \geq$ 42 
erg s$^{-1}$) present in our sample are X-ray obscured.

A soft excess component is required at $>$ 95\% confidence in 4 sources. The origin of
these upturns in the soft portion of the spectrum appears not to be of  the same kind in all sources: in one case (the Seyfert 2 galaxy) 
we parameterize this excess with a power law model as expected for a scattered component, while in the remaining ones a thermal model (i.e. blackbody or Raymond-Smith  model) provides a better fit.    

Finally we have also checked in each source for the possible presence of an emission lines (i.e. those from iron, expected 
at 6.4-6.9 keV rest-frame). We found no significant detection, with typical upper limits on the equivalent width of $\leq$ 350 eV 

\section{Further work}
The present program will continue as new EPIC fields become public. We will also use the observations discussed in Cappi et al. and Foschini et al. (these proceedings) to increase our sample.

Our final goal is to extend this spectroscopic study to $>$ 100 sources and, if possible, to fainter fluxes.
\\

\begin{acknowledgements}

We thank the HELLAS2XMM Team for the optical identifications. 
E.P. thanks Matteo Guainazzi, Andrea De Luca and Alessandro Baldi for helpful discussions on data reduction procedures. 
E.P. gratefully acknowledges the support from MURST for the Program of Promotion for Young Scientists P.G.R.99.
This research has also made use of the Simbad database, operated by the Centre de Donnees Astronomiques de Strasbourg (CDS). 

\end{acknowledgements}

\end{document}